\documentclass[10pt,conference]{IEEEtran}
\input{preamble}
\begin{document}

% \title{A Bug is Being Born. A Time Sensitive Forecasting and Recognizing of the Early Symptoms}

\title{A Defect is Being Born: How Close Are We?\\A Time Sensitive Forecasting Approach}

\author{
\IEEEauthorblockN{
Mikel Robredo\IEEEauthorrefmark{1}, 
Matteo Esposito\IEEEauthorrefmark{1}, 
Fabio Palomba\IEEEauthorrefmark{2},
Rafael Peñaloza\IEEEauthorrefmark{3},
Valentina Lenarduzzi\IEEEauthorrefmark{4}\IEEEauthorrefmark{1},
}

\IEEEauthorblockA{
\IEEEauthorrefmark{1}University of Oulu ---
\IEEEauthorrefmark{2}University of Salerno  ---
\IEEEauthorrefmark{3}University of Milano-Bicocca ---
\IEEEauthorrefmark{4}University of Southern Denmark
}

\IEEEauthorblockA{
\{mikel.robredo, matteo.esposito\}@oulu.fi; fpalomba@unisa.it;  rafael.penalozanyssen@unimib.it; lenarduzzi@imada.sdu.dk
}
}

% \author{\IEEEauthorblockN{Matteo Esposito}
% \IEEEauthorblockA{\textit{University of Oulu}\\
% Oulu, Finland\\matteo.esposito@oulu.fi}
% \and
% \IEEEauthorblockN{Mikel Robredo}
% \IEEEauthorblockA{\textit{University of Oulu}\\
% Oulu, Finland\\mikel.robredo@oulu.fi}
% \and
% \IEEEauthorblockN{Fabio Palomba}
% \IEEEauthorblockA{\textit{University of Salerno}\\
% Salerno, Italy\\fpalomba@unisa.it}
% \and
% \IEEEauthorblockN{Rafael Peñaloza}
% \IEEEauthorblockA{\textit{University of Milano-Bicocca}\\
% Milano, Italy\\rafael.penalozanyssen@unimib.it}
% \and
% \IEEEauthorblockN{Valentina Lenarduzzi}
% \IEEEauthorblockA{\textit{University of Oulu}\\
% Oulu, Finland\\valentina.lenarduzzi@oulu.fi}
% }

% \date{Received: date / Accepted: date}
\maketitle

%\shorttitle{A Bug is Being Born}

% \author{Mikel Robredo}
%    \affiliation{
%    \institution{University of Oulu\\} 
%    \city{Oulu}
%    \country{Finland}
%  }
%  \email{mikel.robredomanero@oulu.fi}

%  \author{Rafael Pe\~naloza}
%  \affiliation{
%    \institution{University of Milano-Bicocca\\} 
%    \city{Milano}
%    \country{Italy}
%  }
%  \email{rafael.penaloza@unimib.it}

%  \author{Valentina Lenarduzzi}
%  \affiliation{
%    \institution{University of Oulu\\} 
%    \city{Oulu}
%    \country{Finland}
%  }
%  \email{valentina.lenarduzzi@oulu.fi}

% \renewcommand{\shortauthors}{Robredo, et al.}

\maketitle

\begin{abstract}
    \textit{Background.} Defect prediction has been a highly active topic among researchers in the Empirical Software Engineering field. Previous literature has successfully achieved the most accurate prediction of an incoming fault and identifying the features and anomalies that precede it through just-in-time prediction. As software systems evolve continuously, there is a growing need for time sensitive methods capable of forecasting defects before they manifest.
    \textit{Aim.} Our study seeks to explore the effectiveness of time sensitive techniques for defect forecasting. Moreover, we aim to investigate the early indicators that precede the occurrence of a defect.
    \textit{Method.} We will train multiple time sensitive forecasting techniques to forecast the future bug density of software project, as well as identify the early symptoms preceding the occurrence of a defect.
    \textit{Expected results.} Our expected results are translated into empirical evidence on the effectiveness of our approach for early estimation of bug proneness.
\end{abstract}

% \begin{abstract}
% Fault prediction has been a highly active topic among researchers in the Empirical Software Engineering field. Its modeling focuses on providing the most accurate prediction of an incoming fault and identifying the features as well as anomalies that precede it. Previous literature has demonstrated valuable insights after conducting fault prediction studies through Machine Learning and Deep Learning techniques. 
% However, \rsc{these techniques do not reach to thoroughly capture the existing temporal dependencies among observations} as a factor to describe the emergence of a bug. This registered report aims at proposing an empirical study design to \rsc{explore the effectiveness of multivariate Time Series Analysis techniques for fault forecasting. In fact, } we aim to investigate the early indicators that precede the occurrence of a bug. In this way, this work strengthens the existing \rsc{current knowledge} of Time Series Analysis as a suitable \rsc{methodology} for the fault forecasting research, and further, it contributes to the body of knowledge within the Empirical Software Engineering field.
% \end{abstract}


%However, the vast majority of these techniques did not consider the time dependence among observations as a factor to describe the emergence of a bug. This registered report aims at proposing an empirical study design to assess the effectiveness of Time Series Analysis techniques in validating the reliability of fault forecasting. Moreover, we aim to investigate the early indicators that precede the occurrence of a bug. In this way, this work strengthens the confirmation of Time Series Analysis as a suitable technique for the fault forecasting research, and further, it contributes to the body of knowledge within the Empirical Software Engineering field.    

% \keywords{Time Series Analysis, Software Maintenance, Empirical Software Engineering }

% \todo[inline,color=blue!50]{RPN: may I suggest for the title: ``Forecasting and Recognizing Early Symptoms''?}

\section{Introduction}
\label{sec:Intro}

% \todo[inline]{FABIO1: We should not compete against JIT defect prediction, since our models and prediction aim for continuous variable prediction}
% \todo[inline]{FABIO2: Therefore, one thing is a classification of  “defect/non-defect” and another is “defect proneness”}
% \todo[inline]{We should look into variables that provide a parallel prediction of continuous variables “how close we are of getting a defect”}

% \todo[inline, color=green]{Mikel@: Intro reviewed. Review again after paper is ready}
Developers frequently modify the source code to introduce new features or rectify defects~\cite{Lehman1980} \ReviewerB{during} the software maintenance and evolution process. However, these modifications may inadvertently introduce new defects~\cite{kim2008classifying}, necessitating careful verification by developers to ensure that such changes do not introduce flaws in the code. This verification task typically occurs either directly during development (e.g., by running test cases) or during code reviews~\cite{Bacchelli2013}. 

An effective strategy for allocating inspection and testing resources to the portions of the source code more likely to be defective is through defect prediction~\cite{Hall2012}. 
Defect prediction involves constructing statistical models to anticipate the defect-proneness of software artifacts, primarily by leveraging information related to the source code or the development process~\cite{pascarella2019}.

Over the past decade, the issue of defect prediction has garnered significant attention from researchers. They have endeavored to tackle this problem through two main approaches: (i) conducting empirical studies to identify factors that contribute to artifacts being more defect-prone and (ii) proposing innovative prediction models designed to accurately forecast the defect-proneness of source code. Recent defect prediction research has tried to address this challenge with multiple approaches based on Machine Learning and Deep Learning algorithms~\cite{zhao2023systematic} and using supervised and unsupervised techniques~\cite{li2020systematic}. These recent studies perform just-in-time (JIT) defect prediction using future binary classification. They chronologically order the training data and therefore perform online prediction based on past historical data~\cite{mcintosh2018fix, tabassum2020investigation, tan2015online}. \ReviewerB{Moreover, prior research has investigated potential model-agnostic explanation techniques for defect prediction~\cite{DBLP:journals/tse/JiarpakdeeTDG22}, as well as explored Deep-Learning enabled approaches to render software defect prediction more transparent and explainable~\cite{DBLP:conf/kbse/KhananLPJTCRS20, DBLP:conf/msr/HoangDK0U19}. Our study aims to follow the same path and investigate the \textit{early symptoms} or software indicators that precede the occurrence of a bug before it happens.}

Although these studies have significantly advanced our understanding of defect prediction, they all focus on binary prediction over the occurrence of a defect. In parallel, already existing time-sensitive forecasting models have demonstrated promising results for probabilistically estimating the occurrence of an event, for instance, Time Series Analysis (TSA) techniques~\cite{robredo2024evaluating}, Bayesian techniques~\cite{harrison1976bayesian}, as well as novel techniques based on the Transformer architecture~\cite{peixeiro2026time}. Consequently, we aim to explore the effectiveness of already existing time-sensitive forecasting approaches to estimate the occurrence of a defect probabilistically, and therefore \textbf{support} the results of already existing JIT prediction techniques with prior estimated knowledge. Furthermore, we aim to investigate the early indicators that precede the occurrence of a defect when employing the considered approaches.

% The majority of current techniques assess the defectiveness of software artifacts through long-term predictions. By analyzing information gathered from prior software releases, these models anticipate which artifacts are more likely to be prone to defects in future releases~\cite{li2020systematic, zhao2023systematic}. 

% However, most prediction techniques do not study the temporal dependencies and trends over time and therefore do not consider the time-dependence factor among observations thoroughly when performing predictions.~\cite{li2020systematic}. More precisely, some prediction models tend to be trained in a batch learning setting, where there is no temporal order between the training instances, and therefore the temporal nature of the historical data is lost~\cite{zhao2023systematic}.

% While the performance reported highlights the suitability of some of the conducted approaches, it is still unclear whether these techniques are the only support for professionals in finding the exact moment when the defect is induced based on historical data.

% In contrast, we believe that Time Series Analysis (TSA) techniques also form a robust approach. They can exclusively identify the temporal dependencies and trends that precede a defect, and consequently forecast its' introduction in the code as well as provide forecasting intervalsconsidering the existing observations and their mentioned temporal dependencies. Further, we believe that there exist potential variables that can show early symptoms that help identify the imminent introduction of a defect.

Therefore, in this registered report, we design an \textbf{exploratory study} design to explore the effectiveness of time-sensitive techniques for defect proneness
estimation, and further inspect which are the early symptoms that precede the introduction of a defect. In this sense, we aim to provide three main contributions to the proposed study:

\begin{itemize}
    \item Supporting the current knowledge of fault/defect prediction by exploring the effectiveness of time-sensitive techniques for defect proneness estimation.
    \item Exploring the impact of different time-windows on the defect-proneness estimation effectiveness.
    \item A detailed identification of the indicators that the most descriptive independent variables in defect forecasting demonstrate early, before and during the introduction of the defect.
\end{itemize}

\noindent\textbf{Paper structure.} Section~\ref{sec:relworks} introduces the background and related work, while Section~\ref{sec:ES} describes the empirical study design. Section~\ref{sec:Threat} identifies some potential threats to validity and Section~\ref{sec:Conclusions} draws the goals of the presented study and highlights some future work.

% We are able to clearly identify when a defect/fault is introduced and fixed  in the source code using SZZ algorithm that label the commit fault-inducing and the commit fault-fixing. We are able to depict “what” (issues, such as code smells, syntactic violations, anti-pattern…) is present in the two categories of commits.

% So, we can derive “what has changed” in the two commits. We can have different possible scenario, such as:
% \begin{itemize}
%     \item scenario 1: in the commit fault-inducing we have only a code smells X and in the commit fault-fixing no code smells (healthy)
%     \item scenario 2: in the commit fault-inducing we have only a code smells X and in the commit fault-fixing the code smell Y
%     \item scenario 3: in the commit fault-inducing we have only a code smells X and Z and in the commit fault-fixing no he code smell Y ….
% \end{itemize}

% Moreover, we can analyze  the “history” of the code before the commit fault-inducing, in order to investigate what “causes” have lead to the fault/defect introduction (\textbf{Why}). We also could derive \textbf{how} the  fault/defect was introduced in order to discover the \textbf{early symptoms} that can provide a sort of “wake-up call” to the developers. 

%\input{Section/Background.tex}
\section{Background and Related Work}
\label{sec:relworks}
In this section, we outline the theoretical background and discuss previous studies related to our empirical study.

% \todo[inline]{Matteo: @Mikel ho rielaborato questa sezione, tagliato, sforbiciato ed introdotto i baesiani e i transformers. Please, have a look at it :)}
\subsection{Software Defect Prediction}
% \todo[inline]{Matteo: @Mikel rimuovi un po di riferimenti da qua, dobbiamo tagliare 11 referenze}
Software defect prediction is a long-standing research area in software engineering, with extensive literature exploring how to identify fault-prone components~\cite{li2020systematic}. Traditional approaches estimate defect proneness at release time, whereas \ReviewerB{JIT} prediction shifts the focus to commit-level granularity~\cite{fan2019impact, mockus2000predicting}. Building on these foundations, continuous defect prediction frameworks~\cite{madeyski2017continuous} and anomaly-based techniques~\cite{neela2017modeling, Afric2020} have enabled large-scale empirical analyses of software evolution.
Both product metrics (e.g., McCabe’s complexity~\cite{mccabe1976complexity}, CK suite~\cite{chidamber1994metrics}) and process metrics (e.g., code churn~\cite{liu2017code}, entropy~\cite{hassan2009predicting}) have proven to be reliable predictors of defect-proneness~\cite{basili1996validation, graves2000predicting, pascarella2020performance}. Subsequent work emphasized the importance of commit-level characteristics such as developer experience, commit size, and code entropy~\cite{SZZ}, showing that large or complex commits are more likely to introduce defects. Recent machine and deep learning models further improved predictive accuracy and interpretability~\cite{kamei2012large, pascarella2019, lomio2022machine, DBLP:conf/msr/HoangDK0U19}, yet they remain inherently reactive, focused on detecting defects after they are introduced rather than understanding when they begin to emerge. \ReviewerB{Consequently, Jiarpakdee et al.~\cite{DBLP:journals/tse/JiarpakdeeTDG22} conducted a large-scale empirical study on model-agnostic explanation techniques in order to explore how instance-level explanation techniques are valuable for understanding why and then software components are predicted as defective. Building on these ideas, Khanan et al.~\cite{DBLP:conf/kbse/KhananLPJTCRS20} proposed JITBot, an explainable JIT defect prediction chatbot enabling not only the defect proneness of software components but also explaining the factors contributing to the occurrence of a bug.}

While traditional defect prediction studies mainly focus on classifying whether a change or file is defective, often at release or commit time, \textbf{our study} shifts the perspective toward a \textit{time-aware} and \textit{evolutionary} understanding of defects. Instead of treating a bug as a binary outcome, we model its temporal progression, aiming to anticipate \textit{when} a defect is likely to emerge. Our temporal perspective enables us to identify the \textit{early symptoms} of defect formation, capturing what emerges before a bug manifests. Hence, our approach provides a continuous probabilistic view of defect birth and evolution, thereby \textbf{overcoming} the static, post-hoc nature of state-of-the-art defect prediction models and offering a significant advantage for proactive quality assurance and preventive maintenance.

\subsection{Retrospective inspection of \ReviewerB{defects life cycle}}
Previous research has investigated the benefits of labelling defective commits based on the Affected Versions (AV) reported in the project's Issue Tracking Systems (ITS)~\cite{SZZ}. \ReviewerA{Since there is no systematic specification of all the AVs within an issue ticket}, \ReviewerB{Vandehei et al~\cite{vandehei2021leveraging} and Falessi et al.~\cite{falessi2021impact} defined the \textit{Stable Proportion} (P) method.} The P method is computed based on the temporal dimension of the evolution of a defect across AVs (see Figure~\ref{fig:bug_detection}). In their proposed method, a defect is first injected in the code base at the \textit{Injection Version} (IV). Subsequently, the creation of a defect report is made within the project's ITS, which can be matched by date with the Opening Version (OV) of the defect. Thus, the stable life cycle of a defect finishes with the defect-fixing commit being registered within the version control history of a project (C$_i$)~\cite{esposito2023uncovering}. Therefore, for each defect, the versions preceding the IV are labelled as not affected, while versions
from the IV to the Fixing Version (FV) are labeled as AV. Thus, in the absence of an IV detailed in an issue ticket, the stable life cycle of a defect can be heuristically estimated by the proportion of the number of
versions required to discover and to fix a defect, that is, $FV-OV$ proportional to $FV-IV$. The authors preferred this adoption to first compute the value of P, and therefore be able to label all the AVs for each detected defect in a software project.

\subsection{Time-Aware Analysis}

\paragraph{Time Series Analysis Approaches} 
\ReviewerB{TSA} provides a statistical foundation for understanding temporally ordered data~\cite{brockwell2002introduction}. These techniques uncover dependencies over time, enabling predictions of future trends from historical patterns. The choice of model depends on key temporal properties such as non-stationarity, seasonality, and external influences~\cite{brockwell2002introduction}.  

In software defect prediction, TSA techniques have been used to capture the temporal dynamics of software quality. Wu et al.~\cite{wu2010time} pioneered this direction by comparing ARIMA, X12-enhanced ARIMA, and polynomial regression on Debian defect data. Their results showed that the X12-enhanced ARIMA model achieved the best forecasting accuracy. Extending this work, Pati and Shukla~\cite{pati2013time} applied univariate autoregressive neural networks and hybrid ARIMA–NN models, demonstrating that hybrid approaches yield the most accurate forecasts.  

Building on these foundations, our study extends TSA applications by incorporating additional independent variables. Through this approach, we aim to identify early temporal indicators of software defects and detect symptoms of potential defects as they emerge during development.

\paragraph{Bayesian approaches}
 Bayesian approaches, in particular, offer a principled way to handle uncertainty and causal dependencies among software metrics. Okutan and Yıldız~\cite{okutan2014software} used Bayesian networks to reveal probabilistic relationships between code metrics and defect proneness, showing that network-based reasoning can improve interpretability and stability. More recently, Kumar et al.~\cite{kumar2025bayesian} applied a Bayesian Belief Network enhanced with feature ranking and CK metrics, achieving an accuracy of 77.9\% in predicting fault-prone modules. Compared with classical classifiers such as Decision Trees or Random Forests, Bayesian models yield more stable results across datasets and allow combining quantitative code measures with qualitative process factors~\cite{sharma2023bayesian}. In our context, where defect prediction operates at commit and file levels, Bayesian inference offers a natural extension, capturing probabilistic dependencies between process metrics (e.g., churn, entropy) and defect outcomes, while explicitly modeling uncertainty. 

\paragraph{Transformer-based approaches}
In parallel, transformer-based architectures have recently shown strong potential for software defect prediction by capturing both syntactic and semantic information from source code. Liu and Zhou~\cite{liu2025dpfusion} introduced DP-TFusion, a transformer model that fuses abstract syntax tree (AST) sequences with metric-based features, significantly improving cross-version prediction performance. Similarly, Han et al.~\cite{han2025transformer} demonstrated that transformer models outperform recurrent and traditional machine learning approaches, especially on large and imbalanced datasets. Furthermore, Zhang et al.~\cite{zhang2023hierarchical} proposed a hierarchical transformer to jointly model token-level and line-level contexts, achieving higher precision at the line granularity. In our study, transformer-based representations could encode the sequential and structural nature of code changes, capturing dependencies across preceding and succeeding statements. 

\section{Empirical Study Design}
\label{sec:ES}
\begin{figure*}[h]
    \centering
    \includegraphics[width=\linewidth]{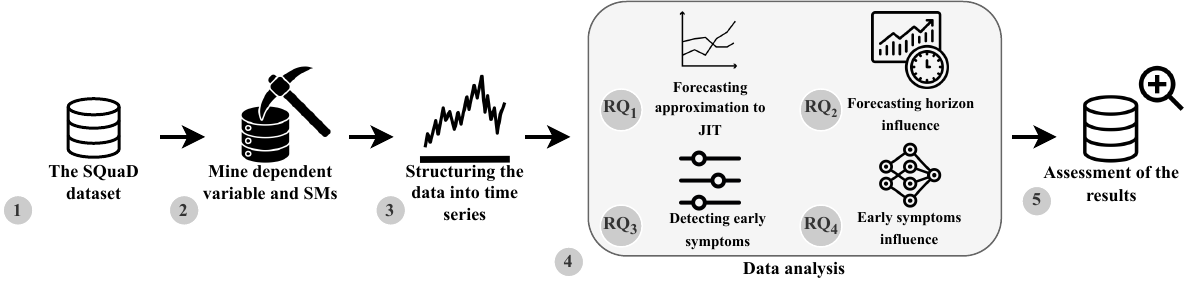}
    %\captionsetup{font=footnotesize}
    \caption{Overview of the study design (SQuaD: Software Quality Dataset)}
    \label{fig:study_diagram}
\end{figure*}

In this section, we describe the designed empirical study. This includes the goal and the research questions, the study context, the data collection, and the data analysis (see Figure~\ref{fig:study_diagram}). 
We design our empirical study based on the guidelines defined by Wohlin et al.~\cite{wohlin_experimentation_2024}. In the designed study, we investigate the effectiveness of defect-forecasting models grounded in \ReviewerB{TSA}~\cite{robredo2024evaluating}, Bayesian inference~\cite{harrison1976bayesian}, and modern Transformer-based architectures~\cite{peixeiro2026time, liu2024timer}. For better readability, we refer to these approaches as \textit{models} throughout the paper. \ReviewerA{Similarly, we refer to all the \textbf{product}, \textbf{process}, and additional \textbf{quality} metrics as Software Metrics (SM) to ease the reading comprehensiveness throughout the paper.}

To allow the replication of our study, we will publish the raw data. 

% \todo[inline, color=green]{@Mikel: Figure: Show difference between JIT and our approach (show how we help JIT), there's some similar figure in one of Matteo's work.}
%\todo[inline, color=green]{@Mikel: Emphasize that JIT is going to be used as actual input}
%\todo[inline, color=green]{@Mikel: Remove all the content related to the Time Series Analysis and refer to our paper for those insights}

\subsection{Goal and Research Questions}

% \todo[inline, color=green]{@Mikel: NEW RESEARCH QUESTIONS}
% \todo[inline, color=green]{@Mikel: RQ1) Expand to different training formats, expand to different forecasting horizons, check which are the best models from the ones employed}
% \todo[inline, color=green]{@Mikel: What are the early indicators for the apparition of a defect?}

% We formalize the goal of this study according to the GQM approach~\cite{Basili1994} as follows:

% \vspace{2mm}
% \begin{tabular}{lp{4.3cm}}
% \textit{Analyze} & time-sensitive defect-proneness forecasting\\
% \textit{for the purpose of} & assessing its effectiveness \\
% \textit{with respect to} & established JIT defect prediction \\
% \textit{from the point of view of} & researchers and practitioners \\
% \textit{in the context of} & open-source software projects
% \end{tabular}
% %
% \vspace{2mm}

The \textbf{goal} of this empirical study is to \textit{analyze} time-sensitive defect-proneness forecasting, \textit{for the purpose of} assessing its capability to support and complement existing defect prediction literature, \textit{with respect to} its effectiveness in providing earlier probabilistic estimates of defect occurrence and in identifying the early indicators (symptoms) that precede defect introduction.
\textit{from the point of }view of researchers and practitioners,
\textit{in the context of} open-source software projects.

Based on this goal, our first Research Question (\textbf{RQ$_1$}), is:

% \todo[inline]{Matteo: @Mikel non saprei se stiamo stimando la defect proneness (quella e una metrica) io direi che stiamo stimando il numero di defect.... ovvero la quantita di defect... no?}
% \todo[inline]{Mikel@Matteo: Vedi mo}
	\begin{boxC}
		\textbf{RQ$_1$.} \emph{To what extent can time-sensitive forecasting models estimate the density of defects of a project over time?}
  \end{boxC}

Prior research has investigated the predictive performance of defect forecasting models~\cite{wu2010time, pati2013time}, showing that \ReviewerB{TSA} approaches can effectively predict future defect counts. The Software Engineering (SE) community has also reported promising results in related forecasting applications~\cite{robredo2024evaluating}.

Our research question focuses on whether defect forecasting models can approximate the injection of a defect within the code base of a project over time. For that, we adopt a \textit{heuristic baseline} for defective commit labelling over the version control  history of a project. We refer to previous research works on defect proneness to define the number of defects per commit~\cite{vandehei2021leveraging, falessi2021impact, esposito2023uncovering}, and therefore use it as \textit{best-effort} ground truth to evaluate the approximation effectiveness of our forecasting outcomes. We evaluate this approximation using standard error metrics such as MAPE, MAE, and RMSE~\cite{robredo2024evaluating}, and we evaluate performance in both one-step-ahead and next-observation horizon settings.

% Accordingly, our research question focuses on whether aggregating JIT binary prediction outputs, yielding the total predicted defects per project version, can be used to probabilistically approximate JIT defect prediction outcomes. We quantify this approximation using standard error metrics such as MAPE, MAE, and RMSE~\cite{jahanshahi2020predicting, robredo2024comparing}, and we evaluate performance in both one-step-ahead and next-observation horizon settings.

Beyond short-term predictions, our study further examines the long-term potential of defect forecasting models to complement current defect prediction knowledge. Specifically, we analyze their effectiveness across multiple future horizons. Hence, we ask:

	\begin{boxC}
		\textbf{RQ$_2$.} \emph{How does the length of the forecasting horizon influence the predictive accuracy and stability of defect forecasts?}
  \end{boxC}	 

A key challenge in time-sensitive defect forecasting lies in producing multi-step forecasts, that is, predictions extending several time steps into the future. Long-term defect forecasts can complement defect prediction by highlighting the sustained probability of defects in a codebase. While single-step forecasts are useful for short-term actions, an area where JIT defect prediction already performs effectively~\cite{pascarella2019, pascarella2020performance, falessi2023enhancing}, multi-step forecasting offers additional value by providing a probabilistic view of how close a system is to future defects. Prior work has demonstrated this potential: \cite{jahanshahi2020predicting} achieved accurate forecasts up to three months ahead, while \cite{robredo2024evaluating} extended forecasting horizons to three years. Such forward-looking analyses can improve developers’ situational awareness and foster more proactive software maintenance practices~\cite{zhao2023systematic, di2017developer}.

Accordingly, in RQ$_2$, we aim to generate long-term forecasts estimating how current code changes may influence future defect occurrences. To achieve this, we evaluate forecasting performance across multiple time horizons using two complementary settings. For each horizon, we compute standard error metrics, MAPE, MAE, and RMSE, to quantify predictive accuracy. Finally, to assess the significance of long-term forecasting effects, we  statistically test the following hypotheses:

\begin{itemize}[leftmargin=1cm]
\item[$H_{1.01}$] \textit{There is no difference in forecasting effectiveness across different models by time horizons.}
\item[$H_{1.11}$] \textit{There is a significant difference in forecasting effectiveness across different models by time horizons.}
\end{itemize}

Moreover, we are keen to investigate whether the best model and the best windows statistically perform better than the others. Hence, we perform a post hoc analysis testing the following hypothesis:
\begin{itemize}[leftmargin=1cm]
    \item[$H_{1.02}$] \textit{There is no significant pairwise difference in forecasting effectiveness between any two models by time horizons.}
\item[$H_{1.12}$] \textit{There is at least one pair of models with a statistically significant difference in forecasting effectiveness by time horizons.}
\end{itemize}

%  This should be put in the data analysis
% Hypothesis testing depends on the distributional assumptions fulfilled by the data~\cite{casella2024statistical}. Hence, following statistical inference practices in existing research literature~\cite{robredo2024analyzing, saarimaki2025does}, we consider testing the normality of the data using \textit{Anderson-Darling} (AD)~\cite{anderson1952asymptotic} and \textit{Shapiro-Wilk} (SW)~\cite{shapiro1965analysis} normality tests. Based on the normality testing, we plan to use parametric tests such as Repeated-Measures \textit{ANOVA}~\cite{hollander1973nonparametric} or non-parametric tests such as \textit{Friedman} test or \textit{Wilconxon} signed-rank test~\cite{casella2024statistical}, for instance.

The forecasted defect-proneness in the long term might be positive or negative, but it will not happen spontaneously. Over time, different SMs demonstrate specific patterns that might explain the development of the defect-proneness probability. Therefore, we ask:

% \begin{comment}
% For that, among the software metrics already studied in former works and considered correlated with a defect presence (see Section~\ref{sec:Variables}) we will identify the software metrics that demonstrate to highly describe the behavior of the defect introduction during the forecasting process, \rsa{that is, which are the "defect early symptoms" that precede the introduction of a new defect}. Furthermore, we will track the changes (positive and negative) in these metrics in order to identify the anomalies that precede the induction of a new defect.     
% \end{comment}

% Our \textbf{hypothesis} ($H_1$) is that behavior indicators such as the \textit{trend} of the highly descriptive software metrics exhibit anomalies prior to and during the introduction of a defect. (\textit{Null Hypothesis} $H_{02}$: The considered software metrics do not exhibit anomalies in their behavior as symptoms of an incoming defect). 

%\todo[inline]{Vale@Mikel: sistema RQ2 please}
%\todo[inline, color=green]{Mikel: Lavoro in corso}

	\begin{boxC}
		\textbf{RQ$_3$.} \emph{What are the early symptoms preceding the occurrence of a defect?}
  \end{boxC}	 

% \todo[inline]{Use the set of metrics that after using feature importance algorithms provide the most information}

% \todo[inline]{RQ4: Ablation experiment (Remove the best performing metrics, run the models with the settings of RQ1 and RQ2, compare the result of the models with that of models trained with the best performing metrics)}

Prior research has shown that software defects arise from multiple interrelated factors that emerge throughout software evolution~\cite{kamei2012large}. Although various studies have identified potential influences on defect prediction from diverse sources~\cite{basili1996validation, pascarella2020performance}, the current state of the art primarily focuses on product and process SM as key determinants~\cite{pascarella2020performance, kamei2012large}. Recent work has further refined the understanding of process SM that most effectively characterizes defect occurrence~\cite{falessi2021impact, falessi2023enhancing}.

With RQ$_3$, we aim to investigate the conditions under which defects are more likely to be introduced based on the forecasting results, thereby extending the existing body of knowledge on defect prediction. Specifically, we quantify the relative importance of SM using established feature importance techniques, including Random Forest (RF)\cite{robredo2024evaluating}, Extreme Gradient Boost (XGB)\cite{robredo2025were}, Correlation Analysis\cite{bakhtin2025network}, and Information Gain Ratio (IGR)\cite{esposito2023uncovering}. \ReviewerA{Furthermore, we operationalize these findings in a two-fold strategy. First, we will adopt professionals' preferences on forecasting \textbf{weekly}, \textbf{bi-weekly} and \textbf{monthly} forecasting window lengths~\cite{robredo2024evaluating}, applicable across TSA, Bayesian, and transformer-based models. Second, we will perform multi-step forecasting, assessing the predictive performance of our models over increasing longer prediction horizons. This will enable practitioners to understand the impact that the anticipation level of long-term prediction can have on current software maintenance decisions.}

Finally, we aim to evaluate the contribution of the most informative SM by evaluating how its removal impacts the performance of the defect forecasting models. Hence, we ask:

\begin{center}	
	\begin{boxC}
		\textbf{RQ$_4$.} \emph{Do the early symptoms preceding a defect contribute to better defect forecasting effectiveness?}
  \end{boxC}	 
\end{center}

To understand how different components influence a model’s performance, researchers often conduct \textit{ablation experiments} in which they remove a component, i.e., a feature, and measure how its absence affects the model’s predictive performance~\cite{sheikholeslami2019ablation}. Over the past year, the use of this method has become increasingly common~\cite{wang2024mrca,cai2025deep}. To strengthen the findings of RQ$_3$, we therefore perform an ablation experiment. Specifically, we aim to evaluate the contribution of the most informative symptoms, that is, the most informative SMs (RQ$_3$), by experimenting using the best model–time window pair identified in RQ$_2$. \ReviewerA{Thereby, we will devise ablation experiments to run on each possible combination of SMs to demonstrate further whether their predictive relevance is reflected in the actual prediction and their absence leads to a decrease in forecasting accuracy.}

Therefore, we conjecture the following \textbf{null} and \textbf{alternative} hypothesis:

\begin{itemize}[leftmargin=1cm]
\item[$H_{4.01}$] \textit{There is no difference in model performance when the most informative symptoms are included among the predictors.}
\item[$H_{4.11}$] \textit{There is a significant difference in model performance when the most informative symptoms are included among the predictors.}
\end{itemize}

% Within the field of SE, \cite{wang2024mrca}, conducted a series of ablation experiments to explore the contribution of different data
% sources and the pruning module to improve the time consumption for causal analysis performance on microservices via multi-modal data. Similarly, \cite{cai2025deep} combined parameter tuning with multiple ablation experiments to investigate diverse combinations of containerization properties on the performance of Deep Learning-based container auto-scaling for cloud native micro-services.

% Specifically, we will investigate the behaviour of software product and process metrics on different future horizons with respect to defect forecasting.

% For that, we will initially consider as indicators the software metrics already studied in former works and considered correlated with a defect presence~\cite{falessi2021impact, madeyski2015process, falessi2023enhancing}, among others. We will identify the software metrics that demonstrate to highly describe the behavior of the defect introduction during the forecasting process, that is, which are the "defect early symptoms" that precede the introduction of a new defect. Furthermore, we will track the changes (positive and negative) in these metrics in order to identify the anomalies that precede the induction of a new defect.

\subsection{Study Context}
\label{sec:Context}

% \todo[inline, color=green]{@Mikel: Same approach as for the MSR dataset, check table in notes for that work, that will be the collection criteria}

% \todo[inline]{Matteo@Mikel: 1) Control if we actually did it in the first colleciton stage of the dataset. 2) If we used the SBOM criteria, then collect it too}

% \todo[inline]{Mikel@All: Can someone check if this ref is correct?}

As context, we plan to consider the list of projects from the Software Quality Dataset (SQuaD)~\cite{robredo2025sqadataset}. \ReviewerB{SQuaD provides longitudinal commit-level metrics, defect labels derived from established methods, and high-quality process/product metrics, making it appropriate for time-sensitive defect-forecasting research.} Within their set of 450 Open-Source Software (OSS) projects they include repositories from sources such as the \textit{Apache Software Foundation} (ASF)~\cite{robredo2024evaluating, esposito2023uncovering}, the \textit{Mozilla}~\cite{lamkanfi2013eclipse} and the \textit{FFMpeg} framework~\cite{liang2024curated}, and the Linux kernel~\cite{liang2024curated}. Moreover, their project selection was performed following a systematic mining criteria (see Table~\ref{tab:mining-criteria}) to ensure the quality of projects mined~\cite{kalliamvakou2014promises}. 

% \todo[inline]{The table with the summary is being cooked in Mahti. If it doesn't finish on time then I'll remove it}

% \todo[inline]{it is ok!}

More details about the projects selected are available here ~\cite{robredo2025sqadataset}. 
% Table~\ref{tab:summary-stats} presents the summary descriptive statistics of the projects considered for the study context.

\begin{table}[t]
\caption{Repository mining criteria.}
\footnotesize
\centering
\begin{tabularx}{\columnwidth}{lX}
\hline
\textbf{Criterion} & \textbf{Condition} \\ 
\hline
Archived or forked & Repository is neither archived nor a fork. \\ 
Last activity & The last activity on the repository is less than six months old. \\ 
Contributors & Repository has at least three contributors. \\ 
Star count & Repository has at least 50 or more stars. \\ 
\hline
\end{tabularx}
\label{tab:mining-criteria}
\end{table}

\subsection{Variables}
\label{sec:Variables}

% \todo[inline, color=green]{PENDING get data for metrics in the related work: https://dl.acm.org/doi/pdf/10.1145/3467895 + https://link.springer.com/article/10.1007/s11219-014-9241-7 + https://link.springer.com/article/10.1007/s10664-022-10261-z}

% \todo[inline]{Dependent Variables, Independent Variables, and Cofounders (if we have)}
The dependent and independent variables considered in this study are described below.

\subsubsection{Dependent variable} 
We consider the \textbf{number of detected defects} in commits from the mined projects as the dependent variable. The nature of the dependent variable is therefore \textit{Continuous}. The variable denotes the current number of existing  defects at the commit level.
%As previously addressed, fault-inducing commits are reported by the SZZ algorithm and stored in the Technical Debt dataset. Thus, the existing data for the dependent variable will be based on fault-inducing commits and commits that did not induce a fault. 

\subsubsection{Independent variables} 

As independent variables, we consider \textbf{process} metrics proven to improve the defect prediction performance~\cite{pascarella2020performance, falessi2023enhancing}. Moreover, we consider 111 \textbf{product} metrics collected by the adopted mining tool \texttt{Understand} \footnote{\url{https://docs.scitools.com/manuals/pdf/metrics.pdf}}. Following recent adoptions by the SE community~\cite{bakhtin2025network}, we rely on the SM collected by Understand to train our models. As mentioned, we refer to them as SMs for better readability.

\subsection{Study Execution}
\label{sec:StudyExecution}

This subsection presents the Execution Plan, including data
collection and analysis strategies.

\begin{figure}[t]
    \centering
    \includegraphics[width=\linewidth]{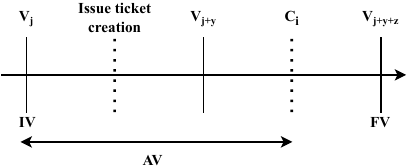}
    \caption{Example of the life-cycle of a defect: Affected Version (AV), Injection Version (IV), defect Fixing Version (FV), defect Fixing commit (C$_i$). (\textit{j, y, z}: Commit observation points)}
    \label{fig:bug_detection}
\end{figure}

\textit{Data Extraction}: This process is composed of three separate tasks as follows:

\begin{itemize}

    \item \textit{Task 1 - Collecting defects through retrospective defect labelling method from Vandehei et al.~\cite{vandehei2021leveraging}} (Figure~\ref{fig:bug_detection}):

    \begin{ReviewerBEnv}
    
    \begin{itemize}
        \item \textbf{Step 1:} We identify the defect-fixing commits (C$_i$) within each project's version control history based on their commit messages~\cite{robredo2024analyzing}.
        \item \textbf{Step 2:} For each defect, we collect from the SQuaD dataset the corresponding issue ticket and its \textit{Injection Version (IV)} when available. Alternatively, we heuristically estimate the IV using \textit{Stable Proportion} P defined by Vandehei et al.~\cite{vandehei2021leveraging}.

        \item \textbf{Step 3:} Using the identified IV and C$_i$, we label all the commits existing within the defect's \textit{Affected Versions (AV)} range. Hence, we collect a list of defects that are active at each commit.

        \item \textbf{Note:} We use commits as temporal anchors; however, our analysis does not focus on individual code changes. Instead, we examine the entire project snapshot at each commit, leveraging the active defect lists to estimate the cumulative defect volume of the project at that specific point in time, namely the commit time. Thus, obtaining a continuous estimation of defect volume.
    \end{itemize}

    \end{ReviewerBEnv}

    \item \textit{Task 2 - Mining process metrics:} We compute the SMs metrics recommended for defect prediction accuracy improvement~\cite{pascarella2020performance, falessi2023enhancing, esposito2023uncovering} by mining the entire change history of the projects.
    
    % According to previous research~\cite{pascarella2020performance, falessi2023enhancing}, we compute the projects SMs defined by Esposito et al.~\cite{esposito2023uncovering} 

    \item \textit{Task 3 - Mining product metrics:} We use \texttt{Understand} commit-wise on each of the considered projects and thus collect values for SMs for their entire change history following already existing approaches~\cite{robredo2025sqadataset}.
\end{itemize}

% \todo[inline, color=green]{@Mikel: The data observations will be commitwise.}
% \todo[inline, color=green]{@Mikel: Metrics will come mostly from Understand + the metrics from https://link.springer.com/article/10.1007/s10664-022-10261-z}
% \todo[inline, color=green]{@Mikel: JIT prediction will be done with WEKA.}

\textit{Data Preprocessing}: Among the models adopted for this study, TSA models require temporally serialized observations for the models to be trained~\cite{brockwell2002introduction}. Since software commits are inherently non-periodic, we serialize the data into regular intervals to use for TSA models, thus generating periodic time-series for each project~\cite{robredo2024evaluating}. Similarly, and following previous research~\cite{robredo2024evaluating}, we employ linear data interpolation for observations resulting from inactive periods without commits. Through this preprocessing step, we aim to ensure consistent and continuous temporal serialization required to run TSA models.

\subsection{Data Analysis} \label{sec:data-analysis}

% \todo[inline, color=green]{@Mikel: Use models from TSA, Bayesian and Transformers}
% \todo[inline, color=green]{@Mikel: For Transformers, check the information from the LR obtained in the Milano course}
% \todo[inline, color=green]{@Mikel: FEASIBILITY ANALYSIS Highlight that we already collected data for release based projects in the MSR work.}

% \begin{figure}[b]
%     \centering
%     \includegraphics[width=\linewidth]{Figure/walkforwardvalidationPOINTS.pdf}
%     \caption{Walk-Forward Optimization. (\textit{i}: New observation.)}
%     \label{fig:walkforward}
% \end{figure}

As this is a registered report, we have not yet executed the study and therefore cannot know \textit{a priori} whether the collected data follow a normal distribution. Hence, we define a data analysis protocol that accounts for both normal and non-normal conditions.  

\paragraph{Forecasting Effectiveness ($RQ_1$ and $RQ_2$).}
To address $RQ_1$ and $RQ_2$, we evaluate the forecasting effectiveness of the three model families, Time Series Analysis (TSA), Bayesian inference, and Transformer-based architectures, using \textit{heuristic defect deduction method} results as the ground truth. For each model–horizon pair, we compute standard error metrics, including MAPE, MAE, and RMSE, to quantify the accuracy of the predicted defect proneness. We train and test our models adopting the \textit{Walk-Forward Optimization} approach~\cite{robredo2024evaluating}. \ReviewerA{Table~\ref{tab:forecasting_models} presents an overview of the specific models used in the study from each of the defined model families. Moreover, and in order to respect the length of registered report publications, we expand the definition of the adopted models as well as the prediction performance evaluation metrics and expected time horizons within the shared online appendix.\footnote{\url{https://doi.org/10.5281/zenodo.17909376}}}

\begin{table*}[t]
\centering
\caption{Overview of the adopted forecasting models.}
\label{tab:forecasting_models}
\renewcommand{\arraystretch}{1.1}
\resizebox{\linewidth}{!}{
\begin{tabular}{lll}
\hline
\textbf{Model Family} & \textbf{Model} & \textbf{Description} \\
\hline
\multirow{4}{*}{\textbf{TSA}} 
 & ARIMA & Univariate TSA model capturing trend and autocorrelation through differencing and autoregressive modelling~\cite{robredo2024evaluating}. \\
 & ARIMAX & Multivariate extension of ARIMA incorporating independent variables~\cite{Mathioudaki2022}. \\
 & SARIMA & Explicit modeling of stochastic seasonal patterns in addition to trend based on the ARIMA model~\cite{robredo2024evaluating}. \\
 & SARIMAX & Seasonal and multivariate extension of the ARIMA model combining independent variables and seasonality patterns~\cite{box1978analysis}. \\
\hline
\multirow{4}{*}{\textbf{Bayesian}} 
 & BDLT & Structural time-series model with damped trend to limit long-term growth~\cite{harvey2013dynamic}. \\
 & BETS & Recency-weighted Bayesian exponential smoothing of level, trend, and seasonality~\cite{hyndman2008admissible}. \\
 & BDLM & State-space model enabling time-varying regression coefficients~\cite{nakajima2013bayesian}. \\
 & BDGLM & Extension of BDLM supporting modeling non-normal distributions of data~\cite{west1985dynamic}. \\
\hline
\multirow{5}{*}{\textbf{Transformers}} 
 & TIMEGPT & Foundation model pretrained on diverse time series~\cite{garza2023timegpt, peixeiro2022time}. \\
 & LAG-LLAMA & Decoder-only probabilistic model for univariate forecasting~\cite{rasul2023lag}. \\
 & CHRONOS & Tokenizes real-valued time series to leverage T5 language-model architectures~\cite{ansari2024chronos, roberts2020exploring}. \\
 & MOIRAI & Probabilistic masked-encoder model using patch-based tokenization~\cite{woo2024unified}. \\
 & TimesFM & Deterministic decoder-only foundation model producing point forecasts~\cite{das2024decoder}. \\
\hline
\end{tabular}
}
\end{table*}

We begin by assessing the distribution of residuals using the Anderson–Darling (AD) test~\cite{andersondarling1952}.  Since software process and forecasting data often deviate from normality, we primarily employ the Wilcoxon signed-rank tests (WT) ~\cite{Wilcoxon1945IndividualMethods} to evaluate differences in forecasting performance across models and time horizons ($H_{1.01}$). WT is a non-parametric test that allows us to compare paired or independent samples without assuming a specific distribution, making them well-suited for our data context. In the case where we fail to accept the \textbf{null hypothesis}, we perform a post hoc comparison using Dunn’s test without a control group ($H_{1.02}$). The Dunn test is a non-parametric post hoc procedure used to identify specific group differences following a significant omnibus result from a rank-based test, such as WT, and extends it to multiple pairwise comparisons by assessing all possible group combinations without requiring a control group. 

Conversely, if the AD test leads us to accept the \textbf{null hypothesis}, i.e., data is normally distributed, we assess the differences in model performance using parametric tests. Specifically, we apply a one-way ANOVA to evaluate whether significant differences exist in forecasting accuracy across models and time horizons. ANOVA compares the means of the performance metrics, MAPE, MAE, and RMSE, across the different model–horizon combinations, using the F-statistic to test the null hypothesis of equal means. If the ANOVA indicates statistically significant differences, we proceed with a Tukey Honest Significant Difference (\textbf{HSD}) post hoc test to identify which pairs of models differ significantly.

% \todo[inline]{Matteo: @Mikel per evitare ambiguita, da qualche parte definisci software metrics = tutte le metriche raccolte, e poi fai semre riferimento a software metrics (SM)}
\paragraph{Most Informative Symptoms ($RQ_3$).}
To explore $RQ_3$, we examine which SMs most strongly influence defect proneness forecasts. We compute feature importance using well-established approaches, including Random Forest (RF)~\cite{robredo2024evaluating}, Extreme Gradient Boosting (XGB)~\cite{robredo2025were}, Correlation Analysis~\cite{bakhtin2025network}, and Information Gain Ratio (IGR)~\cite{esposito2023uncovering}. These methods reveal which SMs most consistently contribute to accurate predictions across models and horizons.  

\paragraph{Ablation Experiment ($RQ_4$)} For $RQ_4$, we conduct an ablation study to quantify the contribution of the most informative SMs. Using the best model–horizon pair identified in $RQ_2$, we compare forecasting performance before and after removing these top-ranked features. Therefore, we test $H_{4.01}$ using WT to detect significant differences in model accuracy, as it effectively captures variations in paired, non-normally distributed samples. Similarly to $RQ_1$, in the case of data normally distributed, we use a paired t-test to compare the forecasting performance of models before and after removing the most informative SMs. The paired t-test evaluates whether the mean difference in performance metrics (e.g., MAPE, RMSE) between the two configurations is significantly different from zero, providing direct evidence of how much predictive accuracy depends on the identified key SMs. If the t-test reveals significant changes, we interpret them as evidence that the removed SMs substantially contribute to model performance. 

In all statistical tests, we adopt a significance level of $\alpha = 0.01$ to maintain a robust balance between Type I and Type II errors~\cite{mishra2019descriptive}.  

% \todo[inline]{Vale, @mikel @matteo: questo pezzo da "For $RQ_4$," non deve andare sotto?}

\subsection{Replicability}
\label{sec:replicability}
To allow the replicability, we publish the raw
data and the code to reproduce our experiments in a replication
package.

% \subsection{Replicability}
% \label{sec:Replicability}

% In order to allow the replication of our study, we will provide the raw data to be used through our online appendix.
% ~\footnote{\label{package} \url{LINK}}.

% \input{Section/Results.tex}
% \input{Section/Discussion.tex}
\section{Threats to Validity}
\label{sec:Threat}

In this section, we discuss the main threats to the validity of our study following the categories defined in empirical software engineering research~\cite{wohlin_experimentation_2024}. 
% \textbf{Construct validity} concerns whether the variables and measures used in the study accurately capture the intended concepts. \textbf{Internal validity} relates to the soundness of the causal relationships inferred from the observed data. \textbf{External validity} addresses the extent to which our results can be generalized to other contexts beyond the studied projects. Finally, \textbf{reliability validity} refers to the consistency and reproducibility of the study process and results. 

\textbf{Construct Validity.} Although the Software Quality dataset provides high-quality repositories, irregular commit activity across projects may result in missing data within the generated time series. We mitigate this threat by applying linear interpolation and data imputation to ensure temporal continuity. Nevertheless, these techniques produce approximations of the real values, and the results might differ if project activity were more consistent over time. Furthermore, we operationalize SMs consistently across all projects to ensure measurement validity and avoid conceptual ambiguity in the constructs under study.

\textbf{Internal Validity.} We consider commits as the main unit of observation since they represent the points at which faults are introduced or detected. The independent variables, process and product metrics, are selected based on their strong empirical association with defect-proneness reported in prior research. Still, we cannot rule out the possibility that other unobserved factors (e.g., team practices, code review dynamics) may influence the results. To mitigate this, we perform an ablation analysis (RQ$_4$) to quantify the contribution of the most informative SMs and validate their causal relevance to model performance.

\textbf{External Validity.} The study context includes mature open-source projects written in different widely used programming languages, such as Java, Python, or C++, for instance, and covering diverse domains, including frameworks, utilities, and core infrastructure systems. Therefore, the results can be generalized to multidisciplinary projects that share these characteristics. However, the findings may not extend to early-stage, proprietary, or low-activity projects, where development patterns and defect dynamics differ substantially.

\textbf{Reliability Validity.} To ensure reproducibility, we extract all SMs and defects automatically using standardized tools and defect-proneness state-of-the-art methods (e.g., \texttt{Understand}, defect-prone commit detecting through the \textit{heuristic defect deduction method}) and document every preprocessing step in our replication package. All scripts, configurations, and statistical analysis pipelines are made publicly available, ensuring that other researchers can independently reproduce and verify our results.
\section{Conclusions}
\label{sec:Conclusions}
This study design pursues two complementary goals. First, it introduces a comprehensive forecasting protocol for defect prediction that integrates different model families, including Time Series Analysis, Bayesian inference, and Transformer-based architectures. Through this, we aim to assess the extent to which temporal and probabilistic patterns in commit activity can explain and anticipate fault occurrence. Second, the study investigates the influence of SMs on defect proneness, identifying early indicators that precede defect introduction and quantifying their impact through feature importance and ablation analyses. Together, these objectives provide a structured foundation for understanding how software evolution dynamics and SMs interactions shape fault emergence over time.

% As future work, we plan to benchmark the proposed forecasting approach against state-of-the-art Machine Learning and Deep Learning models commonly used in defect prediction research. This comparison will deepen our understanding of the relative strengths of traditional and modern forecasting paradigms and guide the design of more interpretable and adaptive fault prediction systems.

\bibliographystyle{IEEEtran}
\bibliography{main}
% \printbibliography

\end{document}